\documentclass{jpp}
\usepackage{graphicx}
\usepackage{multirow}

\usepackage[utf8]{inputenc}
\usepackage[T1]{fontenc}
\usepackage{amsmath}
\usepackage{xcolor}

\newcommand{\Pinj}{$P_\text{inj}$}
\newcommand{\Lleg}{$L_\text{leg}$}

\usepackage{xifthen}
\usepackage{xspace}
\newcommand {\mm}[1] {\ifmmode{#1}\else{\mbox{\(#1\)}}\fi}
\def \x {\mm{\mathbf{x}}\xspace}
\def \y {\mm{\mathbf{y}}\xspace}
\def \z {\mm{\mathbf{z}}\xspace}
\def \yp {\mm{\mathbf{y'}}\xspace}
\def \zp {\mm{\mathbf{z_f}}\xspace}

\newcommand{\para}[1]{\vspace{0.5em}\paragraph{\textbf{#1}}}

\usepackage{hyperref}
\usepackage[normalem]{ulem}

\shorttitle{Data-driven model for divertor plasma detachment prediction}
\shortauthor{Zhu et.al.,}

\title{Data-driven model for divertor plasma detachment prediction}

\author{Ben Zhu
  \corresp{\email{zhu12@llnl.gov}},
  Menglong Zhao, Harsh Bhatia, Xue-qiao Xu, Peer-Timo Bremer, William Meyer, Nami Li, \and Thomas Rognlien}

\affiliation{Lawrence Livermore National Laboratory, Livermore, CA 94550, USA}

\begin{document}

\maketitle

\begin{abstract}
We present a fast and accurate data-driven surrogate model for divertor plasma detachment prediction leveraging the latent feature space concept in machine learning research. Our approach involves constructing and training two neural networks. An autoencoder that finds a proper latent space representation (LSR) of plasma state by compressing the multi-modal diagnostic measurements, and a forward model using multi-layer perception (MLP) that projects a set of plasma control parameters to its corresponding LSR. By combining the forward model and the decoder network from autoencoder, this new data-driven surrogate model is able to predict a consistent set of diagnostic measurements based on a few plasma control parameters.
In order to ensure that the crucial detachment physics is correctly captured, highly efficient 1D UEDGE model is used to generate training and validation data in this study. Benchmark between the data-driven surrogate model and UEDGE simulations shows that our surrogate model is capable to provide accurate detachment prediction (usually within a few percent relative error margin) but with at least four orders of magnitude speed-up, indicating that performance-wise, it has the potential to facilitate integrated tokamak design and plasma control.
Comparing to the widely used two-point model and/or two-point model formatting, the new data-driven model features additional detachment front prediction and can be easily extended to incorporate richer physics.
This study demonstrates that the complicated divertor and scrape-off-layer plasma state has a low-dimensional representation in latent space. Understanding plasma dynamics in latent space and utilizing this knowledge could open a new path for plasma control in magnetic fusion energy research.
\end{abstract}

\section{Introduction}\label{sec:intro}
One of the critical concerns for a magnetic fusion power plant is the intense plasma heat flux $q_\parallel$ that flows very rapidly along magnetic field lines to localized material divertor plates. For unmitigated conditions, this heat flux can exceed $10~GW/m^2$\citep{kuang2020divertor} based on extrapolations from current experimental databases and challenges even the most highly developed modern materials with thermo-mechanical engineering limits at around $10~MW/m^2$. So far, the most successful method of reducing this peak heat flux in tokamaks is achieved by \textit{detachment} - that is to use neutral gas to buffer the incoming plasma heat flux and radiate much of the incoming power over a much wider area of the surrounding walls as the recombined neutrals and the radiation photons freely stream across the magnetic field rather than being focused to the plate as for the plasma. In other words, upstream high temperature plasma is extinguished by high density neutral gas before it strikes the divertor plates; hence, it is \textit{detached}.

Because detachment phenomenon involves nonlinear coupled atomic, molecular and plasma physics, it is challenging to obtain accurate and fast detachment predictions (i.e., predict divertor plasma quantities such as density and temperature for prescribed upstream plasma condition). On one hand, theory-based models with simplifications such as the semi-analytical two-point model formatting (2PMF)~\citep{stangeby2018basic} derives analytical formula to calculate divertor plasma density and temperature. These over-simplified 0D models are fast and valuable in terms of providing physics insights, but the corresponding prediction is somewhat crude and relies on \textit{ad-hoc} or data-fitted momentum and energy loss fraction coefficients.
On the other hand, sophistic tokamak edge transport codes, such as 2D UEDGE~\citep{rognlien1994} and SOLPS-ITER~\citep{wiesen2015new} that incorporate a higher level of complexity and physical effects could give fairly accurate predictions but require a few days to a few months to finish one run.
Of course, with a large enough neutral gas and/or impurity injection, detachment almost always occurs; however, ``too much'' detachment can lead to disruption – a rapid termination of plasma confinement, which could also damage the machine.
Thus, predicting, establishing and maintaining a proper degree of plasma detachment, along with a stable and properly positioned ``detachment front'', is a non-trivial requirement. To make this challenge ever greater, current detachment control in existing tokamaks is accomplished only with the aid of \textit{in situ} diagnostics such as infrared camera~\citep{maurizio2017divertor} or Langmuir probe~\citep{eldon2022enhancement} that would not survive in a reactor environment despite that the success of future magnetic fusion reactors, such as ITER, rely on detachment to mitigate excessive heat load on divertor targets.
Therefore, attaining a reliable and fast detachment prediction model and control algorithm is imperative to the magnetic fusion energy research.

To address this challenge, we propose to build a detachment prediction model with the data-driven machine learning approach. 
For the last few years, machine learning methods have gained a lot of interest in magnetized plasma research, as summarized in latest review paper by \cite{anirudh20222022}. Various neural networks have been explored and applied to assorted applications of experimental and computational fusion plasma study, e.g., to accelerate data processing~\citep{van2018fast}, to interpret critical plasma parameter directly from diagnostics~\citep{samuell2021measuring}, to aid plasma shaping via magnetic field control~\citep{degrave2022magnetic}, to forecast disruption~\citep{kates2019predicting,montes2019machine,rea2020progress}, to speed-up nonlinear Fokker-Planck-Landau collision operator in particle simulations~\citep{miller2021encoder}, to apply closures in fluid models~\citep{ma2020machine,maulik2020neural,wang2020deep}, to discover governing physics-based partial observations~\citep{mathews2021uncovering}, and so on. The growing applications of machine learning techniques on plasma physics indicates that data-driven science could provide an alternative approach to scientific challenges.

In this paper, we report a fast yet accurate model-based data-driven detachment prediction model. The rest of the paper is organized as follows. The methodology of our approach is discussed in Section~\ref{sec:method}. Section~\ref{sec:phys} describes the 1D UEDGE model used in the study and outlines the data generation process. Section~\ref{sec:ml} presents the architecture of neural networks and hyper-parameters used in this study. The results of our trained models are disseminated in Section~\ref{sec:results}. Section~\ref{sec:comp} compares 1D UEDGE simulations and our predictive model with the basic two-point model and two-point model formatting. Application of our model, its limitations and future work are briefly discussed in section~\ref{sec:app}. Finally, section~\ref{sec:conc} summarizes the paper.

\section{Methodology}\label{sec:method}

\begin{figure}
  \centering
  \includegraphics[width=0.6\textwidth]{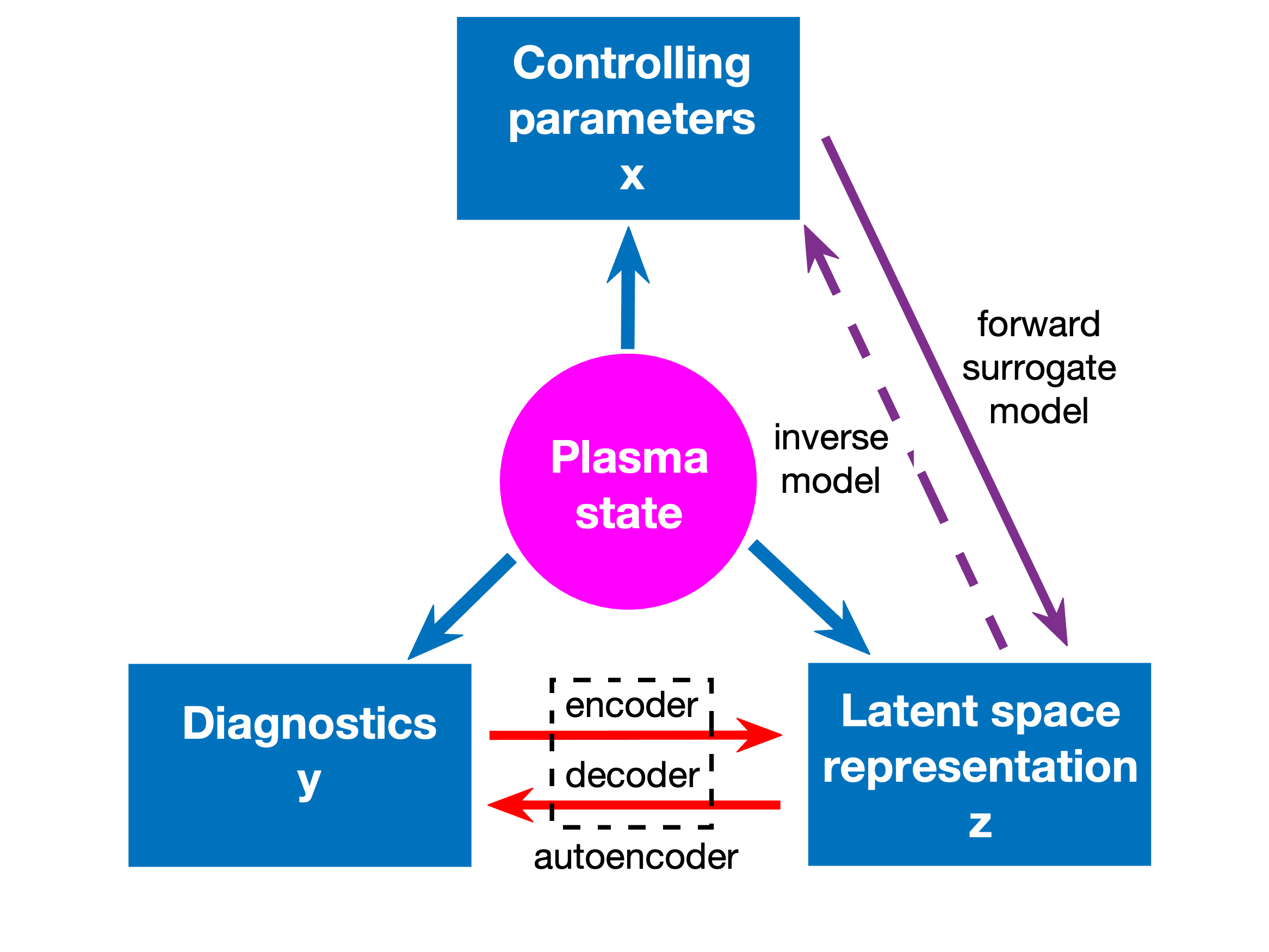}
  \caption{Three descriptions of plasma state.}
\label{fig:plasma}
\end{figure}

Steady-state plasma states are described in two different ways: (1) by control parameters $\x$, or (2) by diagnostic measurements $\y$. For examples, if we neglect any bifurcation or hysteresis phenomena for now, a tokamak plasma is determined by a complete set of engineering and discharge parameters or can be characterized with of diagnosed plasma quantities in detail. Analogously, in numerical simulations, a set of model inputs with proper boundary conditions yields a plasma state that can also be diagnosed synthetically.
A large portion of magnetized plasma research in fact is to discover and exploit the relation between these two descriptions, i.e., predicting laboratory and fusion plasma behavior under certain discharge parameters or finding an optimal control setting for a desired plasma state. In most situations, this is not a trivial task because of the strongly nonlinear nature of magnetized plasmas. 
On one hand, sophisticated numerical models are developed to better understand the physics which governs the complex plasma system, and hopefully to achieve true first principle-based predictive capability.
On the other hand, empirical and simplified models are proposed to provide tools for daily machine operations, experiment planning and future device design.
Often there is a gap in between them - empirical and simplified models are fast but not necessarily accurate, while sophisticated numerical models can provide more reliable predictions but too slow to be used for practical application purpose. 
A fast yet accurate surrogate model can bridge this gap. In this paper, we develop such a surrogate model for divertor plasma detachment prediction with data-driven approach.

Unlike most data-driven surrogate models that directly connect two states $\x$ and $\y$, we take the indirect approach proposed by \citep{anirudh2020improved}. As shown in \autoref{fig:plasma}, in addition to two conventional descriptions of plasma, here we propose a third description --- that in a latent space --- termed as latent space representation (LSR). Latent space, sometimes also referred to as feature space, is a widely used concept in machine learning research in which items with similar features are positioned closely. We first find the LSR, $\z$, of plasma by compressing diagnostics (e.g., synthetic diagnostics mimicking physical diagnostics, such as Langmuir probe, Thomson scattering, bolometer/radiation measurement, etc.) through an autoencoder (described in \ref{sec:ml1}); then forward and inverse models can be trained to relate control parameters $\x$ (engineering or numerical model input parameters such as heating power, gas puffing rate/upstream density, etc) and $\z$. Hence, instead of constructing neural networks $\x \to \z$ or $\z\to\x$, our forward and inverse predictions are $\x\to\z\to\y$ and $\y\to\z\to\x$.
Comparing to the direct approach, the study~\citep{anirudh2020improved} shows that indirect approach has several advantages, including improved predictive performance, and more data efficient. 
In this paper, we focus on forward prediction (i.e., $\x\to\z\to\y$), whereas the inverse prediction (optimization problem) is subject to future work.

\section{Physics model and data}\label{sec:phys}
In this section, we briefly describe the tokamak edge model used to simulate divertor detachment and outline the data generation process.

\subsection{1D UEDGE model}
UEDGE (\cite{rognlien1994,rognlien2002edge}) is a finite-volume simulation code to capture plasma and neutral transport. UEDGE has been used extensively to model the edge region of present-day and future tokamak devices for almost three decades. It has been utilized to study the divertor detachment physics on many machines, such as Alcator C-Mod (\cite{wising1997simulation}), DIII-D (\cite{porter1996simulation}), ITER (\cite{wising1996simulation}), SPARC (\cite{ballinger2021simulation}). UEDGE solves multi-species fluid transport model for plasma and neutral gas, including important atomic physics, such as line radiation from ionization, excitation, and recombination processes in realistic tokamak equilibria. With the fully implicit Newton-Krylov scheme and adaptive stepping to advance all model equations in time, UEDGE is able to quickly evolve the system to a steady state. 

\begin{figure}
  \centering
  \includegraphics[width=0.48\textwidth]{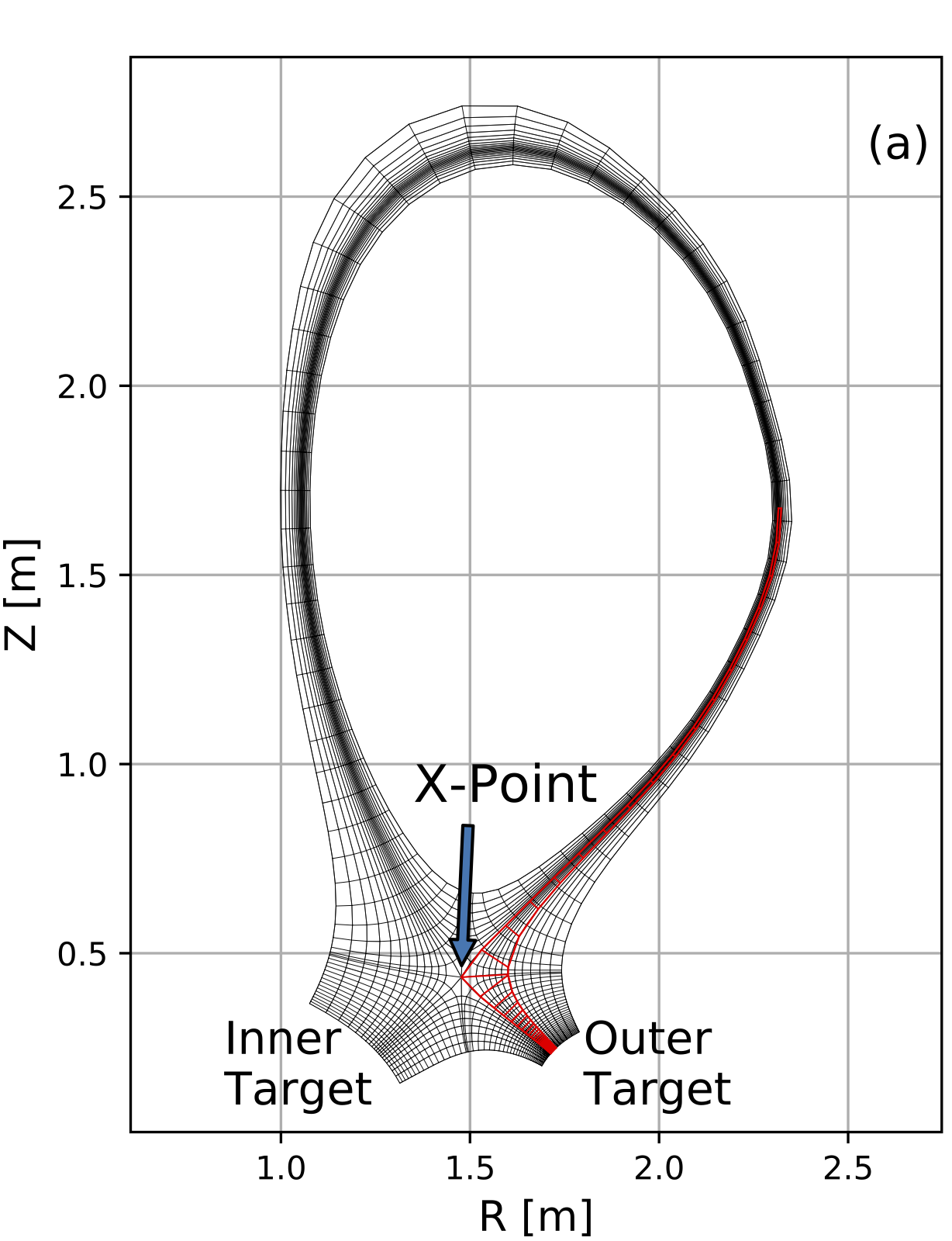}
  \includegraphics[width=0.48\textwidth]{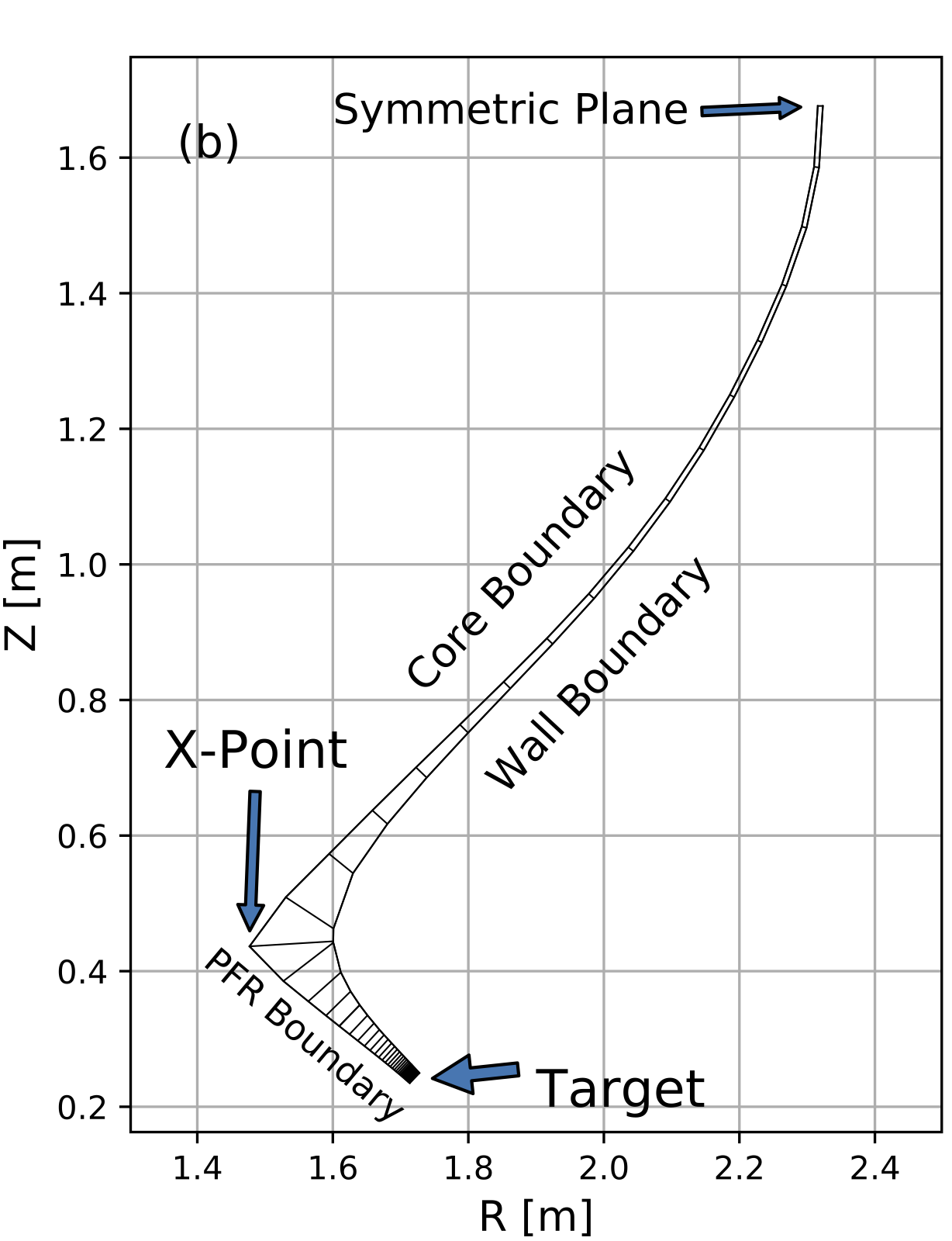}
  \caption{(a) 2D (black) v.s. 1D (red) UEDGE simulation meshes, and (b) sketch of 1D UEDGE simulation setup.}
\label{fig:uedge_mesh}
\end{figure}

In this proof-of-principle study, highly flexible and efficient 1D UEDGE code is used to quickly generate a large training data set that covers a wide range of parameter space beyond normal tokamak operation.
Unlike the commonly used 2D UEDGE model, which simulates the entire tokamak periphery, including closed flux surface region, scrape-off-layer (SOL), and private flux surface region, the 1D UEDGE in this study models only one flux-tube in the low-field side SOL, starting from the outer mid-plane to the outer divertor target plate, as depicted in \autoref{fig:uedge_mesh}(a).
Nevertheless, 1D UEDGE code solves the same set of equations as its 2D counterpart except radial transport and drifts. Plasmas are allowed to transport in the parallel direction that is along the magnetic field whereas neutrals transport poloidally along the flux-tube. As a result, the essence of plasma detachment physics is retained despite the one flux-tube assumption. In fact, simplified 1D geometry disentangles many secondary effects and is sometimes preferred to better quantify the role of different physics processes played in the detachment (\cite{dudson2019role}).
In 1D UEDGE setup, spatial variation of the magnetic field $\boldsymbol{B}$ is retained, i.e. the flux expansion along a flux tube in the SOL is considered in the 1D UEDGE mesh. Power, or the energy influx, across the core boundary is injected at the outer midplane (upstream), and is equally partitioned between electrons and ions. Plasma particles are also assumed to be fueled at upstream. As plasma moves downside the flux tube, it recombines to neutrals which eventually are recycled at the divertor target plate. In this study, the particle and momentum recycling coefficients are set to be $0.99$.

\begin{figure}
  \centering
  \includegraphics[width=\textwidth]{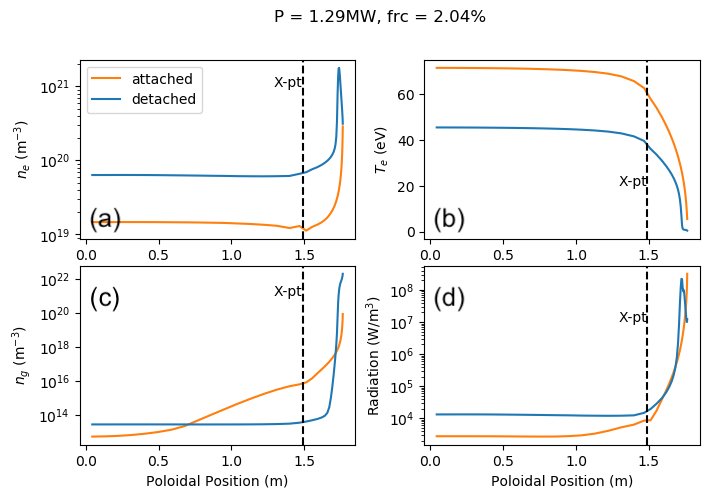}
  \caption{Examples of (a) plasma density, (b) electron temperature, (c) neutral density, and (d) radiation profile of attached (blue) and detached (orange) divertor plasma from 1D UEDGE simulations.
  }  
\label{fig:uedge_1d}
\end{figure}

\begin{figure}
  \centering
  \includegraphics[width=\textwidth]{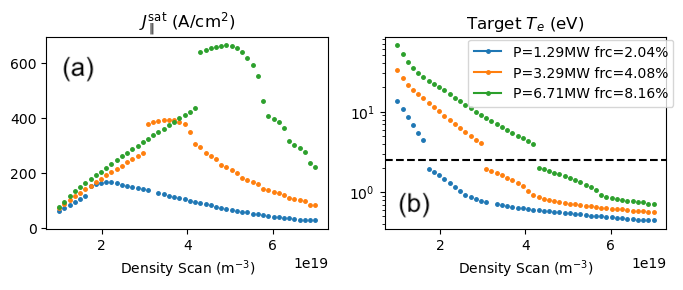}
  \caption{Reproductions of (a) $J_\parallel^\text{sat}$ rollover, and (b) $T_e$ cliff features in 1D UEDGE density scan simulations.}  
\label{fig:uedge_nscan}
\end{figure}

Figure~\ref{fig:uedge_1d} shows the typical attached (in blue) and detached (in orange) plasma in 1D UEDGE simulations. Compared with attached plasma, detached plasma normally has a lower electron temperature $T_e$, a higher neutral gas density $n_g$ at the divertor target, and a distinct detachment front, or radiation peak away from the divertor target.
Not surprisingly, 1D UEDGE code is able to qualitatively reproduce the target ion saturation current density $J_\parallel^\text{sat}$ rollover (e.g., Figure 2 in \citet{loarte1998plasma}) and electron temperature cliff (e.g., Figure 5(a) in \citet{mclean2015electron}) phenomena observed in detachment experiments as shown in Figure~\ref{fig:uedge_nscan}. We admit that more sophisticated models such as 2D UEDGE and SOLPS-ITER considering additional details in principle yield more trustworthy results; but the fundamental physics picture of plasma detachment - that is plasma radiates enough power to reach a low-enough temperature (a few $eV$) and to recombine into neutral gas will not change. Rather than developing a more comprehensive data-driven model with training data from state-of-the-art numerical simulations, here we focus on exploring the data-driven approach for detachment prediction with a somewhat simplified physics-based model (1D UEDGE) that could cover lots of plasma discharge scenarios relatively cheap computationally.

\subsection{Data generation}\label{subsec:data}
Although there are more than one dozen independent geometry and plasma parameters can be adjusted in 1D UEDGE model, here we only vary three most relevant parameters, namely, upstream density $n_{e,u}$, injection power \Pinj~and impurity fraction $f_Z$. These three parameters $(n_{e,u},P_\text{inj},f_z$) will be referred as ``model inputs'' thereafter.

In this study, we use DIII-D tokamak geometry with a fixed divertor leg length \Lleg~$=0.2696~m$, defined as the poloidal distance between the $X$-point to the outer divertor target as shown in Figure\ref{fig:uedge_mesh}(b). We uniformly sample $n_{e,u}\in[1,7]\times 10^{19}~m^{-3}$, \Pinj~$\in[1,10]~MW$ and $f_Z\in 0-10\%$. Because the first wall material of DIII-D tokamak is graphite, here the impurity species is set to be carbon. With 60  sample points for $n_{e,u}$\ and $f_Z$, and 40 sample points for \Pinj, 144,000 cases in total are launched. However, not all of these cases can reach a physically feasible steady state solution within a limited simulation time. Hence, only 111,598 converged simulation cases with good power balance relation are accepted as the training data set. Similarly, an independent validation data set with a coarser sampling rate ($50\times 50\times 30$ and 57,655 accepted solutions) is generated. Unless stated otherwise, the neural networks presented in this paper are trained and validated based on the training data set, only the combined forward prediction model are validated with the validation data set.



\begin{table}
  \begin{center}
\def~{\hphantom{0}}
  \begin{tabular}{cccc}
       data set   & $n_\text{total}$ & $n_\text{detach}(\%)$ & $n_\text{attach}(\%)$  \\[3pt]
       training   & 111,598 & 56,047(50.22) & 55,551(49.78) \\
       validation & 57,655  & 28,468(49.38) & 29,187(50.62) \\
  \end{tabular}
  \caption{Numbers of detached and attached cases and their percentages in the training and validation data sets.}
  \label{tab:data_sets}
  \end{center}
\end{table}

To avoid manual intervention of hundreds of thousands simulations individually, all the UEDGE runs are launched and managed by Merlin \citep{peterson2022enabling} - a machine learning workflow framework designed for large scale HPC environment. It takes 4 days with 200 CPUs to get about $170,000$ converged cases. The data generation time could be further shorten if more CPUs are utilized.

For the diagnostic description of plasma state in this 1D setup, five synthetic measurements are taken across the simulation domain, including the upstream electron temperature $T_{e,u}$, ion saturation current density $J_\parallel^\text{sat}$, electron density $n_{e,t}$ and temperature $T_{e,t}$ at divertor target, as well as the radiation profile $P_{rad}$. As illustrated in Figure~\ref{fig:uedge_mesh} (b), $T_{e,u}$ mimics the mid-plane Thompson scattering measurement, $J_\parallel^\text{sat}$ mimics the Langmuir probe measurement, $n_{e,t}$ and $T_{e,t}$ come from either divertor Thompson scattering or Langmuir probe, and $P_{rad}$ is deduced from either bolometer or C-III emission.
Note here $J_\parallel^\text{sat}$, $T_{e,u}$, $n_{e,t}$ and $T_{e,t}$ are scalars, while $P_{rad}$ is a 1D profile. These diagnostic measurements are collected for all 111,598 cases and then normalized for model training.

\begin{figure}
  \centering
  \includegraphics[width=0.6\textwidth]{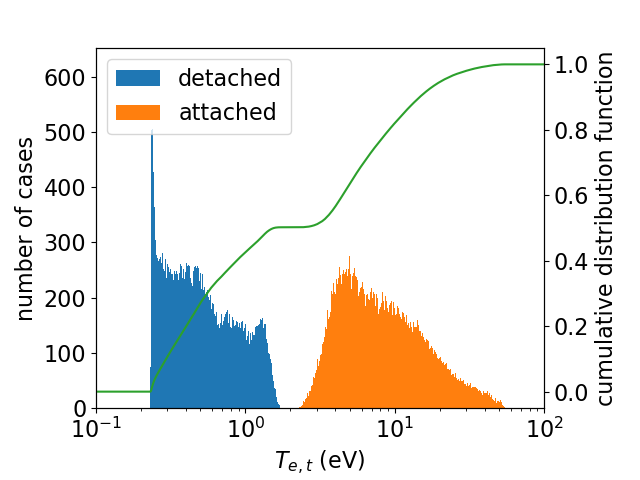}
  \caption{Histogram and cumulative distribution of electron temperature at divertor target $T_{e,t}$ from the training data set (111,598 cases in total).}  
\label{fig:tet_dist}
\end{figure}

Prior model training, all cases are labeled as either ``detached'' or ``attached'' because there is no ``partial detached'' state in the 1D system. Here the choice of detachment criterion is straightforward. 
As shown in Figure~\ref{fig:tet_dist}, electron temperature at the divertor target $T_{e,t}$ has two distinct distributions around $T_{e,t}=2.1~eV$. The gap at $2.1~eV$ is caused by the temperature cliff phenomenon showed in Figure~\ref{fig:uedge_nscan}(b).
Therefore, 56,047 cases with $T_{e,t}<2.1~eV$ are labeled ``detached''; while the rest 55,551 cases are ``attached'' cases. As shown in Table~\ref{tab:data_sets}, nearly half of the cases in the validation data set are labeled ``detached'', similar to the percentage of ``detached'' cases in the training set.
In DIII-D experiment, the temperature cliff occurs around $5~eV$, this apparent discrepancy is mainly due to our simplified 1D setup. The temperature cliff in our simulations is consistent with the sonic transition point where the ion parallel velocity reaches sonic moving away from the target. A thorough investigation of temperature cliff in 1D is beyond the scope of this paper and will be presented in a separated paper by~\citet{zhao2022}.
We would like to point out that labeling cases is primarily for better quantify models' accuracy in Section~\ref{sec:results}; it doesn't affect the training process at all. Also, the lower $T_{e,t}\sim 0.22~eV$ limit shown in Figure~\ref{fig:tet_dist} is likely related to the atomic physics. When $T_e<1~eV$, the ionization rate decreases while the recombination rate increases dramatically. In this temperature range, electron collisionality is very high such that three-body recombination becomes dominant. This process moves plasma potential energy to electron thermal energy, stopping $T_e$ decreasing further.

\section{Development of Machine Learning Models}\label{sec:ml}

\begin{figure}
  \centering
  \includegraphics[width=0.8\textwidth]{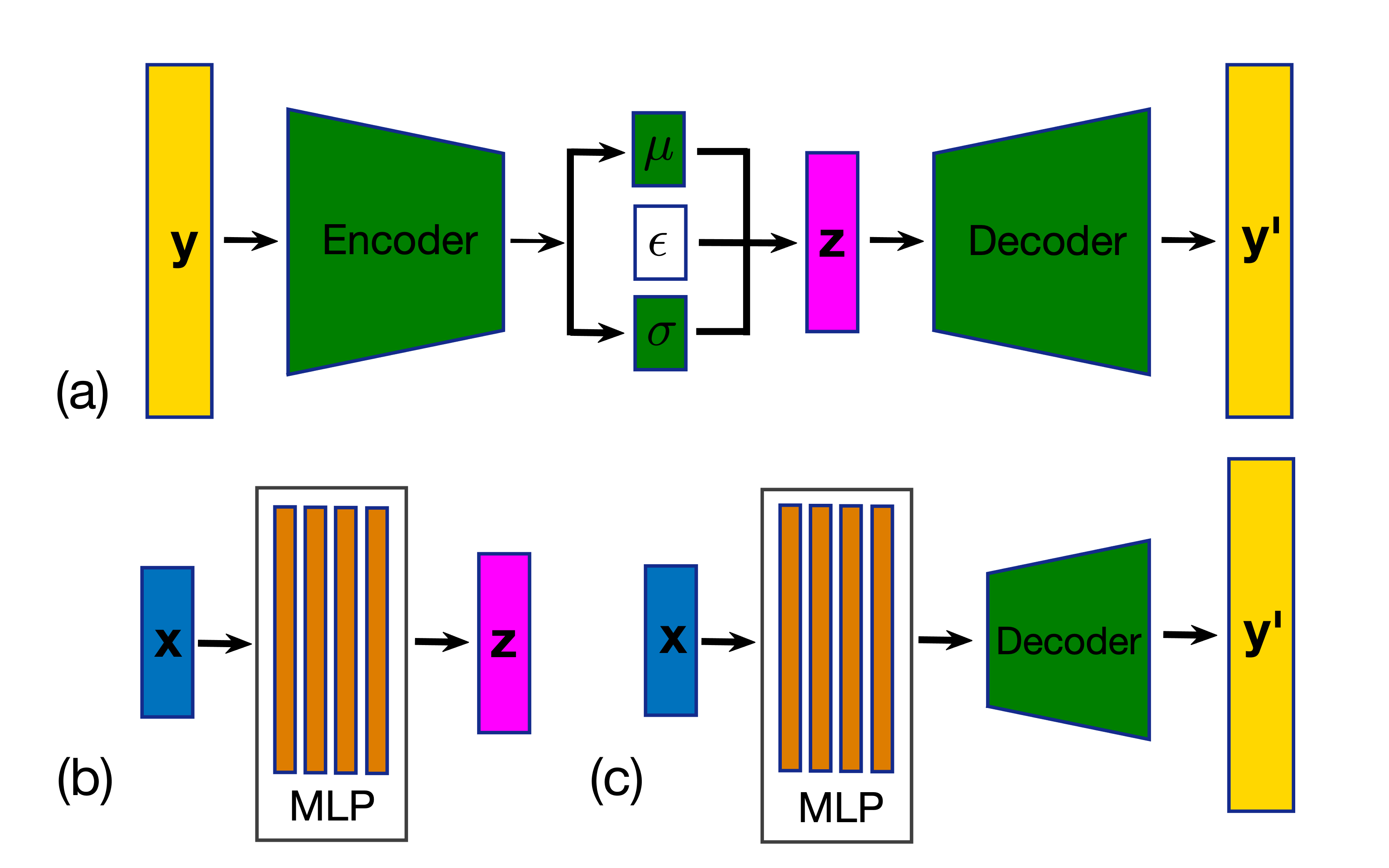}
  \caption{Sketches of (a) $\beta$-variational autoencoder, (b) forward multilayer perceptron (MLP), and (c) combined data-driven model.}
\label{fig:nns}
\end{figure}

The goal of our ML-based approach is to predict the outputs of UEDGE simulations directly from a set of given model inputs, without actually performing UEDGE simulations and with significant speed up. As has been shown in literature~\cite{anirudh2020improved}, this task becomes substantially easier and more tractable if the prediction is made through an intermediate and reduced dimensional space also learned through ML (see \autoref{fig:plasma} and \autoref{fig:nns}). Therefore, in this work, we utilize two ML models. The first ML model --- a \emph{variational autoencoder} --- is used to generate an invertible mapping from ODs to a reduced-dimensional \emph{``latent space''}, whereas the second ML model --- a \emph{surrogate model} --- is used to predict UEDGE outputs in this latent space, given the inputs to UEDGE.

\subsection{Latent Space Identification using Autoencoders}\label{sec:ml1}

The basic idea behind latent space identification (also referred to as dimensionality reduction or data compression) is straightforward. Given an input data point, the goal of an autoencoder (AE) is to reduce it to fewer dimensions such that the reduced representation allows reconstructing the original input as faithfully as possible.
An AE comprises two parts: (1) an encoder that learns to meaningfully encode the data into a set of values --- a compact representation of the input (i.e., latent coordinates), and (2) a decoder that attempts to reconstruct the original inputs from this latent space encoding. By forcing the reconstruction of data, an AE attempts to learn a suitable latent space that captures all necessary degrees of freedom while discarding trivial variations, noise, and redundant correlations in the data.

In this work, we utilize two types of autoencoders to create a suitable latent space representation in two steps. First, we develop a (vanilla) AE to identify an appropriate level of reduction, i.e., the dimensionality of the latent space, and a suitable neural network architecture. Next, we turn this autoencoder into a special type of autoencoder, called the $\beta$-variational autoencoder (VAE), to create the final latent space representation.

\para{AE design to identify the dimensionality of a suitable latent space.}
Given a set of OD measurements (\y), we develop an AE to identify the compressed latent space representation (\z) of plasma state. We use \yp to denote the reconstucted input from the AE. In other words, the encoder maps $\y \to \z$, and the decoder maps $\z \to \yp$. 
In our case, \y (and hence also \yp) has 34 values --- it comprises four scalar values ($J_\parallel^\text{sat}$, $T_\text{e,u}$, $N_\text{t}$, and $T_\text{e,t}$) and a 30-element 1D array ($P_\text{rad}$). However, the different values (scalars vs.\ the array) exhibit different ranges of values. Therefore, we utilize a scaling factor $\alpha_i$ for each of the 34 input dimensions to prevent overfitting on the 30 values for the array. In particular, we choose $\alpha_i = 0.1$ for values corresponding to the array elements and $\alpha_i = 1.0$ for the scalars, effectively asking the model to consider each array element 10$\times$ less important than the scalars. 
Formally, our AE is trained to minimize the loss function
\begin{equation}
    \mathcal{L}(\y,\yp) = \frac{1}{N}\sum_i \alpha_i|y_i - y'_i|^n,
    \label{eq:lossae}
\end{equation}
where $N=34$ is the total number of elements in the input OD data \y. For $n=1$, the loss function is the mean absolute error (MAE) or $L1$ norm, and for $n=2$, it becomes the mean square error (MSE) or $L2$ norm.

Since our input data is simply a collection of values, our AE design uses a series of fully-connected neural network layers. Each such layer provides dense connections between input and output values through a linear transformation (a matrix multiplication with an additive bias) followed by a nonlinear activation. Given a chosen input/output dimensionality and an activation function, the training of the model then learns the appropriate matrix and bias to perform this transformation. We experimented with different depths of the neural network (i.e., the number of such fully-connected layers) and the corresponding dimensionalities. 
All models were trained using the Tensorflow~\citep{tensorflow2015-whitepaper} and Keras~\citep{chollet2015keras} frameworks with the Adam optimizer~\citep{kingma2014adam}. A dataset of 111,598 samples was used for developing the AE, with a randomly selected 80\% subset of this data  for training, and the remaining 20\% for validating the model (i.e., assess its accuracy).

The final AE model, which provided the best reconstruction accuracy (see \autoref{tab:ae_val}), contains three fully-connected layers that progressively reduce the data dimensionality as $34 \to 18 \to 10 \to 6$. Each layer is followed by the sigmoid activation function. The output of the last layer is the desired latent space representation; in other words, a 6-dimensional latent space is adequate to compress a 34-element diagnostic data in our case so as to recover the input with sufficient accuracy.

\para{VAE design to identify a suitable latent space.}
The AE setup described above is capable of generating compressed encoding of the input data. However, such AEs provide no control of the distribution of the data in the learned latent space representation. Typically, this is not a big issue if the AE is used to encode/decide the data, i.e., use the AE (both encoder and decoder) as one model.
However, it is not suitable for our application as our goal is to utilize only the decoder and in conjunction with a forward, surrogate model. Specifically in our case, it is important to provide smoothness guarantees on the latent space representation, otherwise even small prediction errors from the forward model could be amplified significantly by the decoder.
To regularize the distribution in the latent space and to improve decoder's generative capability, we use a specific type of AE, called $\beta$-variational autoencoder (VAE) \citep{higgins2016beta}. 
The key difference between the (vanilla) AE and the VAE is that the VAE forces the distribution of \z to a given reference distribution --- typically, a multivariate normal distribution --- offering some important mathematical guarantees, such as smoothness. To ensure that $\z$ mimics the reference distribution, a regularization term is added to the loss function to measure how different the two distributions are. Specifically, the loss function from \autoref{eq:lossae} is modified to
\begin{equation}
    \mathcal{L}(\y,\yp) = \frac{1}{N}\sum_i \alpha_i|y_i - y'_i|^n \quad + \quad
    \beta D_{KL}(f(\z)~||~f_0),
\end{equation}
where $D_{KL}(f(\z)~||~f_0)$ is the Kullback-Leibler divergence, which measures how the distribution $f(\z)$ differs from the reference distribution $f_0$, and a hyperparameter $\beta$ is used to balance the decoder reconstruction accuracy (the first term in the loss function) and the orthogonality of latent space coordinates (the second term). As is common for VAE, our reference distribution $f_0$ is a multivariate normal distribution, $\mathcal{N}(0,1)$. In many applications, $\beta$ is set to be larger than unity for better separation of independent latent space coordinate; in our case, forcing LSR distribution function to match normal distribution is not necessary. Instead, a small $\beta=10^{-9}$ is found to be suitable.
The VAE used in this study has the same network architecture and training parameters as the AE described above.

\subsection{Predicting Simulation Outputs using a Surrogate Model}

Given the latent space representation, \z, generated by the VAE described above, we next develop another ML model that can predict \z based on the physics model inputs, \x (i.e., global discharge parameters). Due to a small numbers of model inputs (i.e., three varying parameters in 1D UEDGE model) and outputs (i.e., six-dimensional latent space), here a simple yet robust multilayer perception (MLP) is chosen to be our forward model. This MLP has four fully-connected layers with 24 neurons on each layer, uses the rectified linear (ReLu) activation function, and MSE as the loss function. Adam optimizer with a learning rate $0.001$ is used for training the model.

\section{Data-driven model performance}\label{sec:results}

In this section, we discuss the results on the performance and validation of the ML models developed using the methods discussed in \autoref{sec:ml}.

\subsection{Latent Space Identification using Autoencoders}

Through the AE described above, we identify the dimensionality of a suitable latent space, $D_\z$, to be 6. \autoref{tab:ae_val} indicates that with both $L1$ or $L2$ norm, the validation loss (in normalized units) increases as $D_\z$ decreases. The $L2$ norm appears to have a critical turning point at $D_\z=6$; the validation loss increases substantially when $D_\z$ is set to be less than $6$, while for $D_\z>6$, the improvement is less profound.

\begin{table}
  \begin{center}
\def~{\hphantom{0}}
  \begin{tabular}{c|ccccc}
      $D_\z$ & 8 & 7 & 6 & 5 & 4 \\[3pt]
      $\mathcal{L}(\y,\yp)^{n=1}$ & $3.5\times10^{-4}$ & $4.2\times10^{-4}$ & $4.7\times10^{-4}$ & $4.8\times10^{-4}$ & $5.2\times10^{-4}$ \\
      $\mathcal{L}(\y,\yp)^{n=2}$ & $2.5\times10^{-6}$ & $3.1\times10^{-6}$ & $3.3\times10^{-6}$ & $4.5\times10^{-6}$ & $5.2\times10^{-6}$ \\
  \end{tabular}
  \caption{To choose a suitable dimensionality, $D_\z$, of the latent space, we study the reconstruction errors posed by differnet dimensionalities. Given the significant jump in the error when going from 6 to 5, $D_\z$ was found to be the appropriate choice.}
   \label{tab:ae_val}
  \end{center}
\end{table}

\begin{figure}
  \centering
  \includegraphics[width=\textwidth]{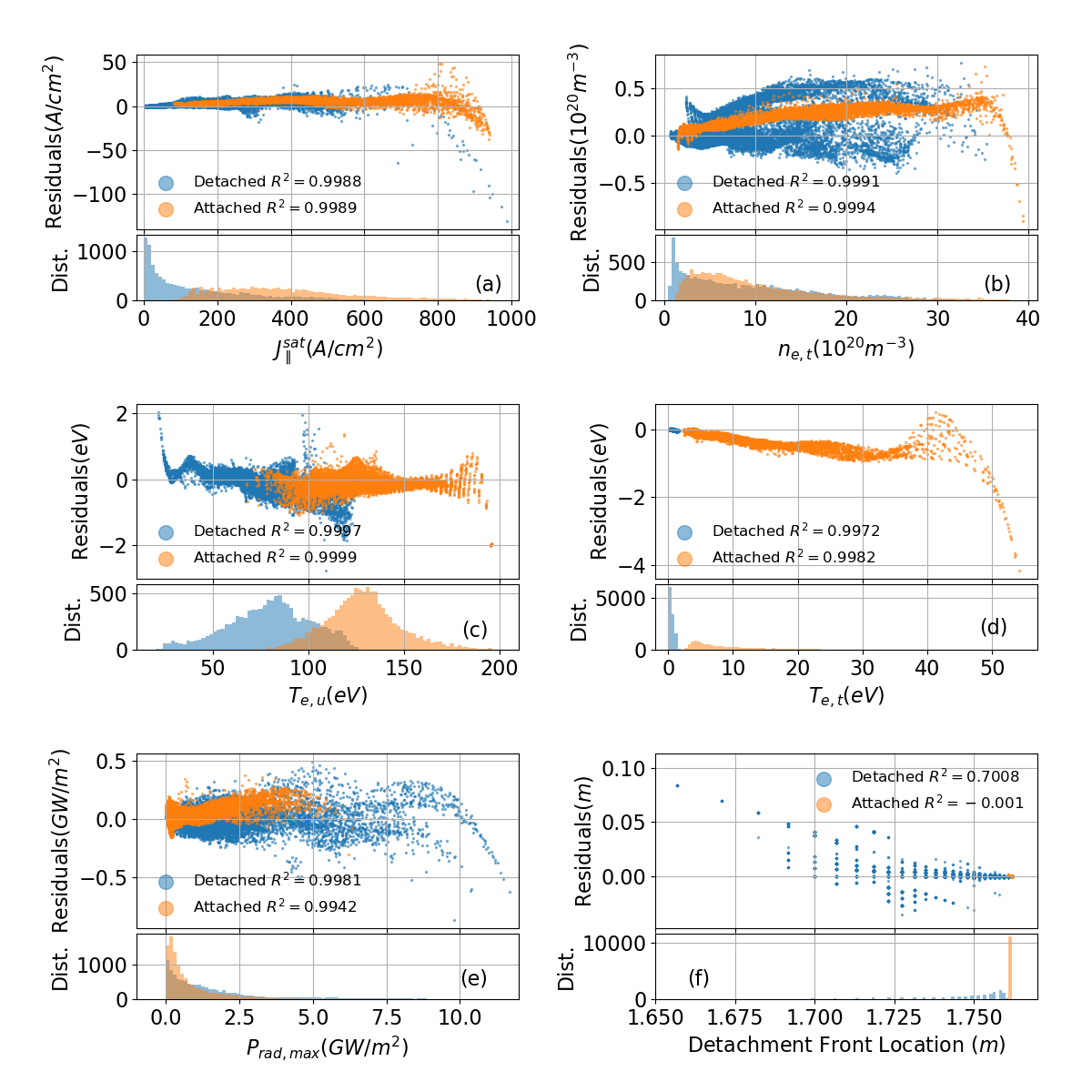}
  \caption{Performance of $\beta$-VAE in terms of absolute residual error for (a) $J_\parallel^\text{sat}$, (b) $n_\text{e,t}$, (c) $T_\text{e,u}$, (d) $T_\text{e,t}$, (e) $P_\text{rad}^\text{max}$, and (f) peak radiation, or detachment front location. Here, $x$-axis is UEDGE (true) value, and the residual error is defined as 
  $\varepsilon=f_{\beta\text{-VAE}}-f_\text{UEDGE}$ for the quantity $f$. 
  All predictions show excellent correlation with the true values, as quantified with almost-perfect $R^2$ scores.
  The distribution of data points are shown in adjoining panels to overcome the issue of overplotting in the scatter plots; these indicate that for most cases, relatively low and unbiased error is obtained, whereas 
  higher errors are associated with relatively smaller number of samples.}
\label{fig:vae_perf}
\end{figure}

Next, we study the performance of the VAE using \autoref{fig:vae_perf} with UEDGE data $f_\text{UEDGE}$ on the horizontal axis and the residual between VAE-reproduced data and UEDGE data $\varepsilon=f_{\beta\text{-VAE}}-f_\text{UEDGE}$ on the vertical axis. If a perfect reconstruction can be achieved, all data points would lie exactly on the horizontal $\varepsilon=0$ lines. In our case, two of the four scalar diagnostic measurements, the electron density at divertor $n_\text{e,t}$ and the electron upstream temperature $T_\text{e,u}$ can be reproduced with excellent accuracy, whereas the other two scalar measurements, the ion saturation current, $J_\parallel^\text{sat}$, and the electron temperature at divertor, $T_\text{e,t}$, also show good quality reconstruction, with a slight performance degradation at high values of the respective diagnostics.
Nevertheless, the $R^2$ scores (or, coefficient of determination that measures the goodness-of-fit between the two variables $f=\{f_i\},g=\{g_i\}$ defined as $R^2=1-\sum_i(f_i-g_i)^2/\sum_i f_i^2$ with $i$ being the index of the variable; hence $R^2=1$ indicates perfect correlation while $R^2=0$ means no correlation) for these four scalars all exceed $0.997$, evidencing a close to perfect replication of input data from trained $\beta$-VAE statistical-wise.
To quantify the reconstruction quality of the radiation profile, $P_\text{rad}$, we choose two metrics: peak amplitude and location. The peak radiation amplitude prediction by the VAE is consistent the UEDGE data with a slightly larger variance ($R^2\simeq0.994<0.997$). The peak radiation, or detachment front location appears to be challenging to be correctly reproduced with $R^2_\text{detached}\approx 0.7$ and $R^2_\text{attached}\approx 0$. However, due to the discrete nature of the simulation mesh, radiation peak locations are discretized as well so that the $R^2$ score here could be misleading, especially for the attached cases. In fact, all 55,551 attached cases whose peaking radiation location at the divertor target are correctly reproduced by $\beta$-VAE but they are represented by what appears to be ``a single'' orange dot in \autoref{fig:vae_perf}(f).

\begin{figure}
  \centering
  \includegraphics[width=\textwidth]{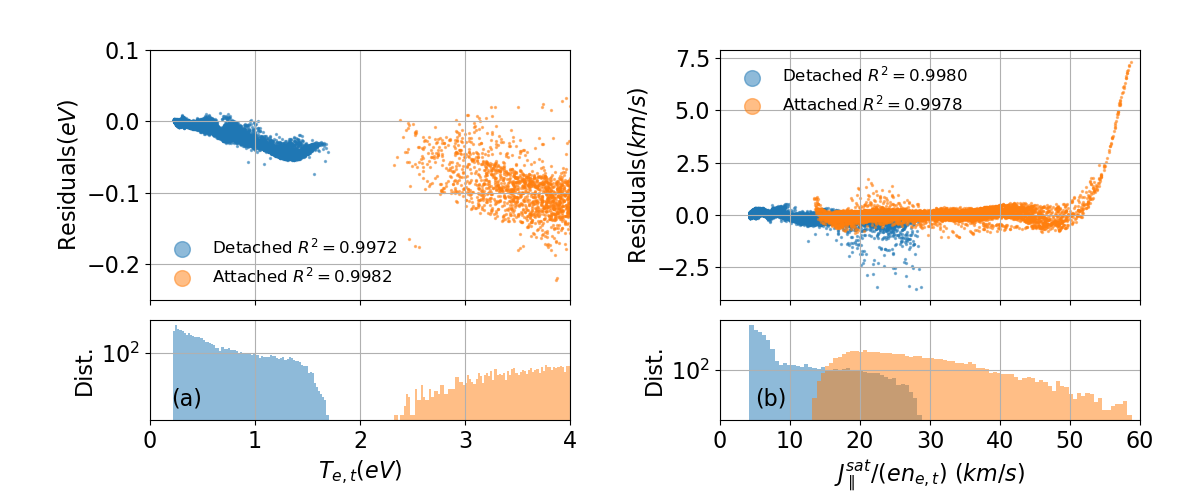}
  \caption{Residual plots for (a) $T_{e,t}$  near detachment transition, and (b) $J_\parallel^\text{sat}/n_{e,t}$. Detached cases are in blue while attached cases are in orange. Note here $y$-axis of the distribution plots is in logarithmic scale }
\label{fig:vae_h}
\end{figure}

One plasma quantity that we are especially interested in for detachment prediction is the electron temperature at the divertor target, $T_\text{e,t}$, as this is arguably the most important and direct indicator of detachment. As shown in \autoref{fig:vae_h}(a), VAE is able to reproduce $T_\text{e,t}$ near the detachment onset value ($\sim2.1~eV$ for this study) very well. \autoref{fig:vae_h}(b) shows that the VAE is able to also accurately retain the ion exist velocity, $v_{\parallel i,e}=J_\parallel^\text{sat}/(e n_\text{e,t})$, which is an inherent quantity that was not trained directly. 
A discrepancy between the VAE's predictions and UEDGE results appears for $J_\parallel^\text{sat}/(en_\text{e,t}) > 55~km/s$, which corresponds to attached plasma cases with $T_\text{e,t}\gg 2.1~eV$. However, this type of discrepancy exists only for less than $2\%$ of the total cases - the small number of $T_\text{e,t}\gg 2.1~eV$ samples may cause the relatively large prediction error for these cases. This result indicates that our trained VAE captures the correct physics constraints between different plasma variables (i.e., autoencoder inputs and outputs) for at least the majority of the cases in our dataset.

\begin{figure}
  \centering
  \includegraphics[width=0.6\textwidth]{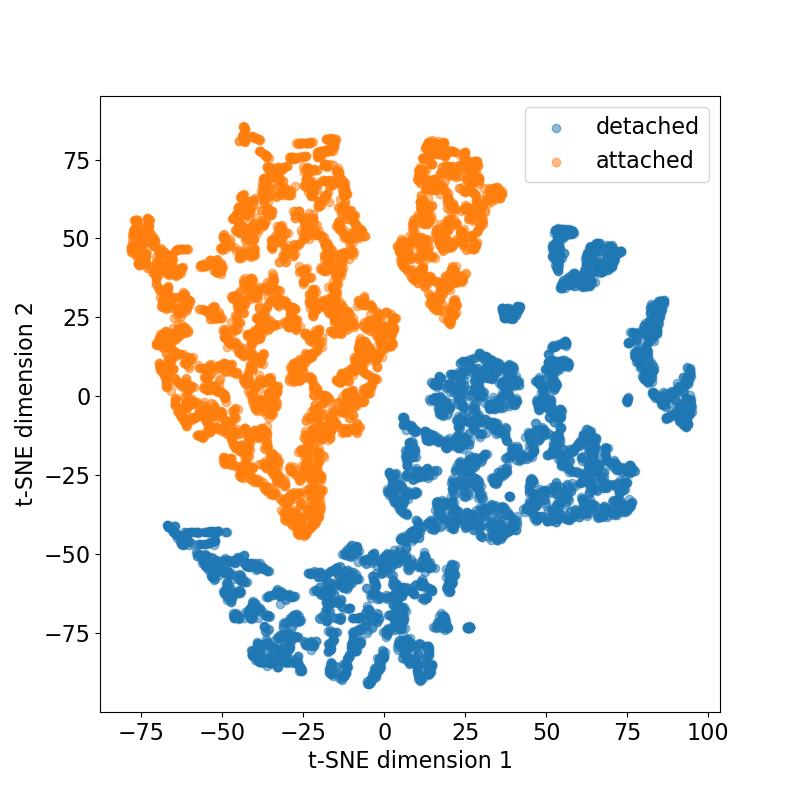}
  \caption{t-SNE visualisation of a subset of 10,000 randomly picked LSR samples in the training data set in which detached (blue) and attached (orange) cases are clearly separated.}
\label{fig:lsr_tsne}
\end{figure}

Finally, we show that $\beta$-VAE encoded LSRs are indeed apart in latent space based on plasma states (e.g., detached or attached). \autoref{fig:lsr_tsne} illustrates the distinctive clusters of LSRs for detached and attached cases in latent space using t-SNE (t-distributed stochastic neighbor embedding) method~\citep{van2008visualizing}. Similarly, Figure~\ref{fig:forward} has the distributions of all six latent coordinates, again separated by attachment/detachment label. The result shows that despite overlaps between detached and attached plasmas in some latent variables (e.g., latent variables 1, 4 and 5), the two scenarios are broadly well separated in the other latent variables (e.g., latent variables 2 and 6), which is a nice feature to have for the upcoming forward detachment prediction.

\subsection{Predicting Simulation Outputs using a Surrogate Model}

\begin{figure}
  \centering
  \includegraphics[width=\textwidth]{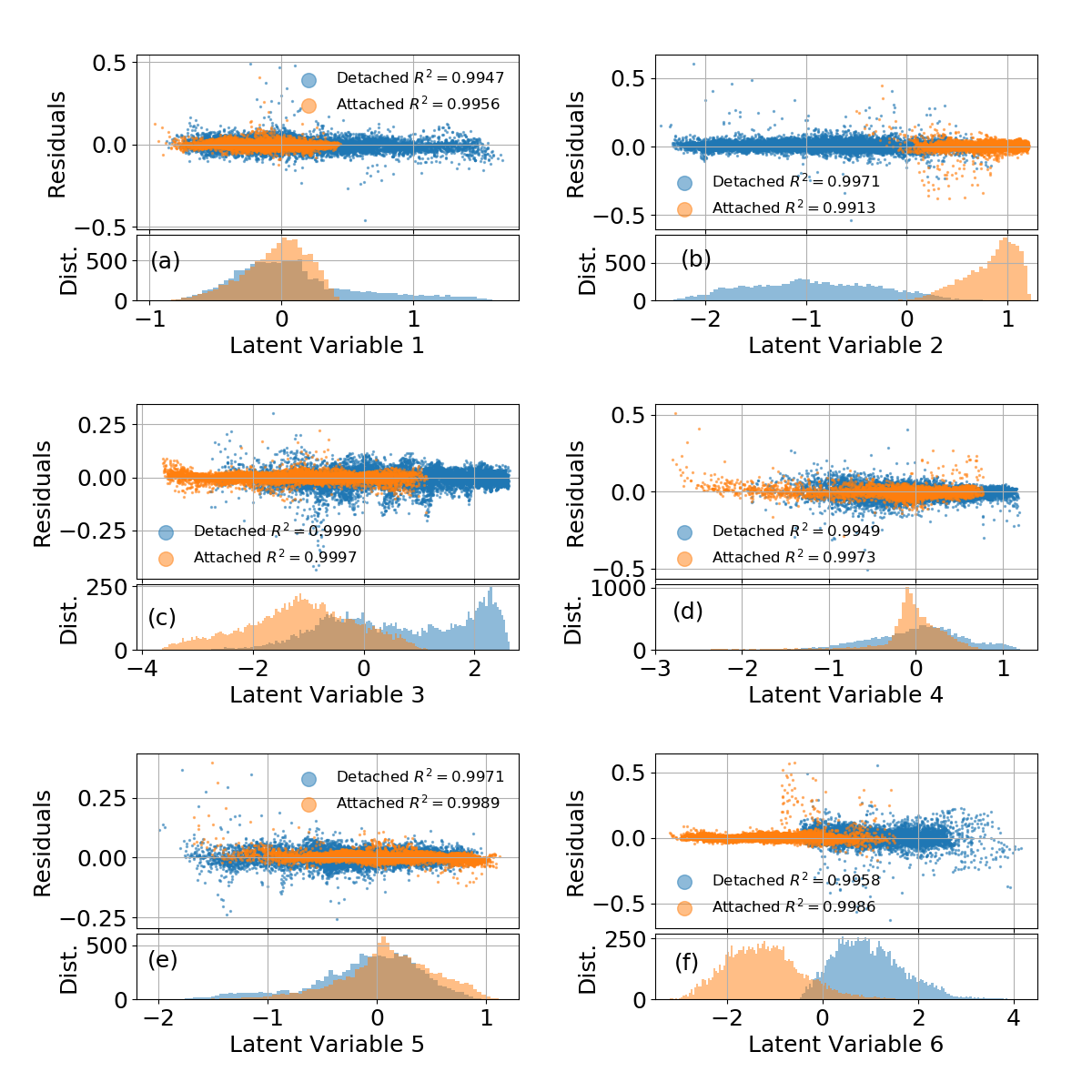}
  \caption{Performance of the MLP model in terms of absolute residual error for all six latent variables. Here $x$-axis is $\beta$-VAE encoded value, while the residual is defined as
  $\varepsilon = f_\text{MLP}-f_{\beta\text{-VAE}}$ for quantity $f$. 
  As with earlier results, an almost-perfect R2 score indicates exceptional prediction quality.}
\label{fig:forward}
\end{figure}

Here, we study the performance of our surrogate model - an  MLP - after 10,000 epochs. In particular, we compare the latent space representation generated by the VAE, \z, and that predicted by the surrogate model, \zp. \autoref{fig:forward} depicts plots the six coordinates within the two quantities (\z vs.\ \zp) as individual scatter plots and shows high degree of correlation. As shown in the figure, the $R^2$ values for all six coordinates exceeds $0.99$, indicating exceptional performance and that our surrogate model is capable of predicting the latent space representation with almost full accuracy.

\subsection{Forward prediction}
With both $\beta$-VAE and forward model trained, we are now able to predict diagnostic measurement of a plasma state from the model inputs by combining the forward model and decoder (Figure~\ref{fig:nns}(c)). To ensure that this model is evaluated properly, a separated validation data set consists 57,655 cases generated independently following the similar data generation process described in Section~\ref{subsec:data}.

\begin{figure}
  \centering
  \includegraphics[width=\textwidth]{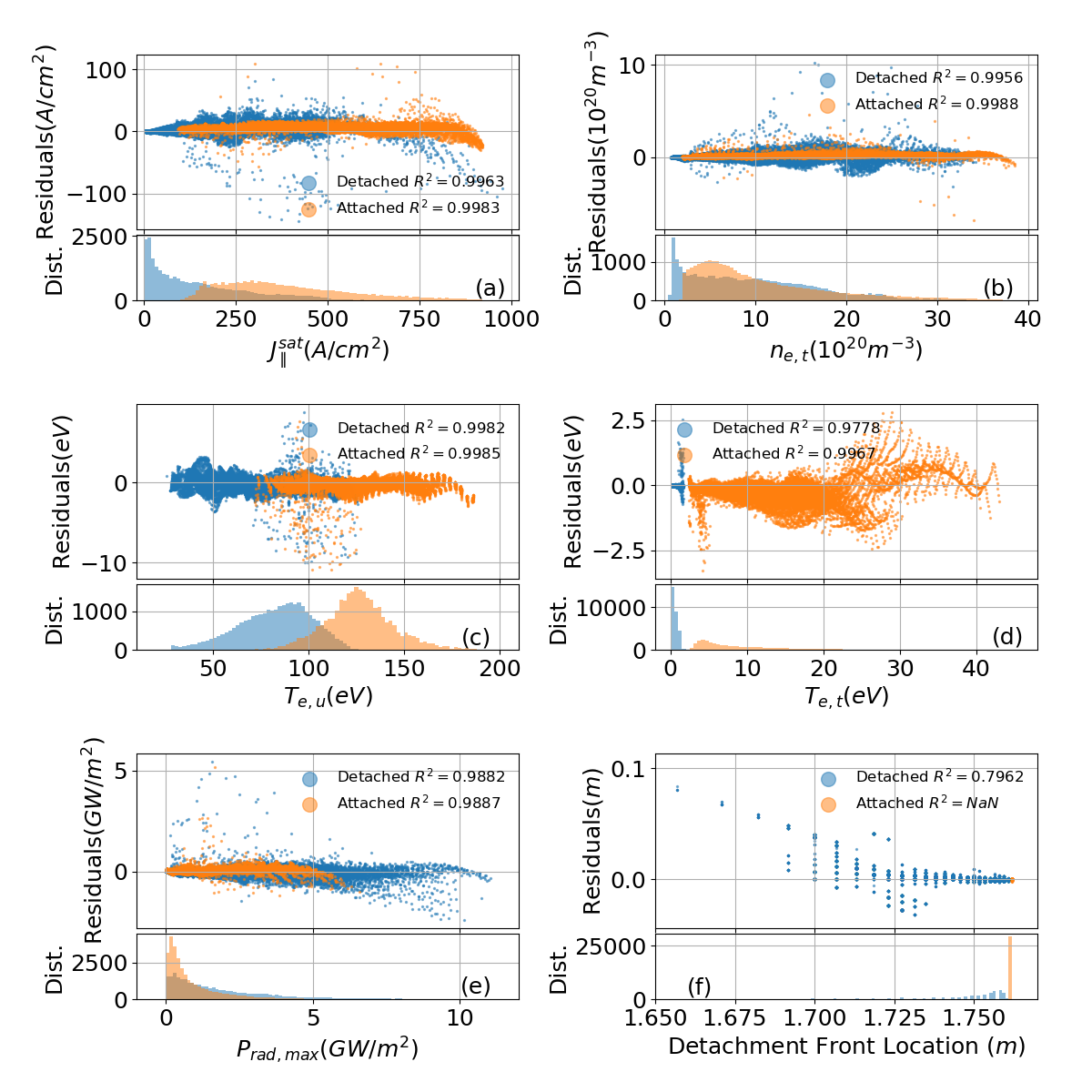}
  \caption{Performance of combined forward prediction model in terms of residuals for (a) $J_\parallel^\text{sat}$, (b) $n_{e,t}$, (c) $T_{e,u}$, (d) $T_{e,t}$, (e) $P_\text{rad}^\text{max}$, and (f) peak radiation, or detachment front location. Here $x$-axis is UEDGE (true) value, while the residual is defined as $f_\text{residual}=f_{ML}-f_\text{UEDGE}$ for quantity $f$. The distribution of $f$ and the $R^2$ scores between $f_{ML}$ and $f_\text{UEDGE}$ for detached (blue) and attached (orange) cases are also provided.}
\label{fig:forwad_pred}
\end{figure}

The performance of forward detachment prediction model is illustrated in Figure~\ref{fig:forwad_pred} where all 57,655 cases are evaluated with the machine learning model then validated with the UEDGE simulation results. As expected, the overall accuracy degrades marginally comparing to the performance of $\beta$-VAE (e.g., Figure~\ref{fig:vae_perf}) in this sequential MLP and decoder architecture. 

\begin{figure}
  \centering
  \includegraphics[width=\textwidth]{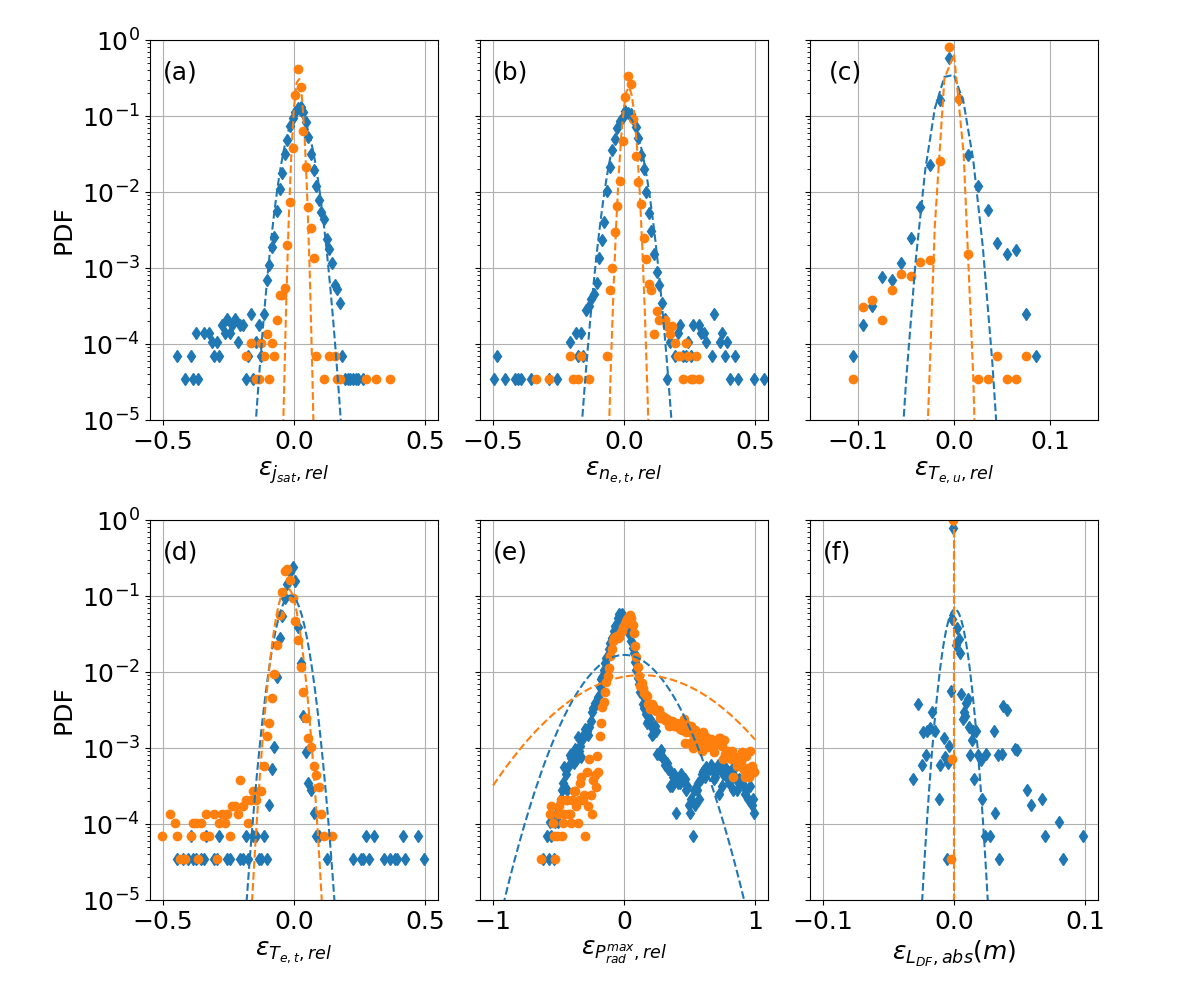}
  \caption{Probability distribution functions (PDFs) of relative error of (a) $J_\parallel^\text{sat}$, (b) $n_{e,t}$, (c) $T_{e,u}$, (d) $T_{e,t}$, (e) $P_\text{rad}^\text{max}$, and (f) absolution error of peak radiation, or detachment front location. Detached cases are in blue while attached cases are in orange. Dashed lines are shifted Gaussian distribution function with $\mu$ and $\sigma$ from Table~\ref{tab:forward_error} correspondingly.}
\label{fig:forwad_error}
\end{figure}

\begin{figure}
  \centering
  \includegraphics[width=\textwidth]{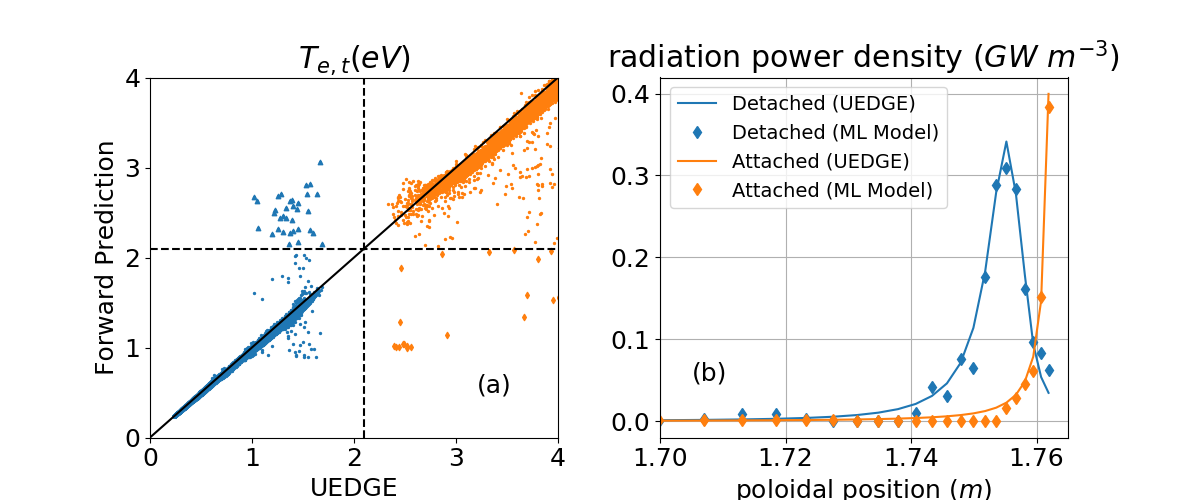}
  \caption{(a) Electron temperature at divertor target $T_{e,t}$ prediction vs UEDGE simulation result, and (b) examples of radiation profile or detachment front prediction. True (UEDGE) detached cases are in blue while attached cases are in orange.}
\label{fig:forwad_highlight}
\end{figure}

Figure~\ref{fig:forwad_error} and Table~\ref{tab:forward_error} elucidate the accuracy of this combined forward detachment prediction model, quantified with error statistic analysis. We examine model accuracy for both detached and attached cases, and find that our model does perform slightly different for different plasma states. For attached plasma, it gives better results when predicting certain plasma quantities, such as ion saturation current and divertor target electron density; while it is less satisfactory for divertor target electron temperature prediction. However, no significant quantitative difference is found except the detachment front location prediction. Out of four scalar measurements, upstream temperature $T_{e,u}$ is the most accurately predicted diagnostic quantity with mean and standard derivation of relative error $\mu=-0.37\%$ and $\sigma=0.83$ for the entire validation data set. The other three scalar measurements are also well predicted with $|\mu|<3\%$ and $\sigma<4\%$ for all the relative errors. The peak radiation strength prediction appears to have a fairly large uncertainty ($\sim10\%$). This is possibly due to the shape peak structure (i.e., lack of resolution) for detached plasmas and boundary effect (i.e., maximum value resides at the last point) for attached plasmas as shown in Figure~\ref{fig:forwad_highlight}(b). Fortunately, the more useful information, the peak radiation location (equivalent to the detachment front location $L_{DF}$ to some extent) prediction is remarkably accurate. $L_{DF}$ is correctly predicted to be at the divertor target plate for nearly all attached plasmas, and has a small uncertainty ($\sigma=0.60~cm$) for detached plasmas. Note that the error distributions of four scalar measurements appear to follow Gaussian distribution, while for the peak radiation amplitude and location, the error distribution are no long Gaussian -- the standard derivation $\sigma$ is skewed by rare (i.e., probability $\sim 0.1\%$) extreme cases. 

\begin{table}
  \begin{center}
\def~{\hphantom{0}}
  \begin{tabular}{|c|c|c|c|c|c|c|}
       & \multicolumn{2}{|c|}{detached} & \multicolumn{2}{|c|}{attached} & \multicolumn{2}{|c|}{combined} \\[3pt] \hline\hline
       & $\mu$ & $\sigma$ & $\mu$ & $\sigma$ & $\mu$ & $\sigma$ \\ \hline
       $\epsilon_{J_\parallel^\text{sat},rel} (\%)$ &  1.66 & 3.76 &  1.67 & 1.26 & 1.67 & 2.79 \\
       $\epsilon_{n_{e,t},rel} (\%)$ &  1.06 & 3.97 & 1.81 & 1.66 & 1.44 & 3.05 \\
       $\epsilon_{T_{e,u},rel} (\%)$ & -0.43 & 1.05 & -0.30 & 0.52 & -0.37 & 0.83 \\
       $\epsilon_{T_{e,t},rel} (\%)$ & -0.13 & 3.90 & -2.70 & 3.06 & -2.01 & 3.57 \\
       $\epsilon_{P_{rad}^{max},rel} (\%)$ & 0.19 & 23.76 & 13.39 & 43.85 & 6.88 & 36.00 \\
       $\epsilon_{L_{DF},abs} (cm)$ & 0.07 & 0.60 & 0 & 0 & 0.04 & 0.42 \\
  \end{tabular}
  \caption{Mean ($\mu$) and standard deviation ($\sigma$) of relative or absolute error of forward prediction model.}
  \label{tab:forward_error}
  \end{center}
\end{table}


We are particularly interested in $T_{e,t}$ prediction as this is the primary indicator of detachment. If we use the same detachment criterion (e.g., $T_{e,t}=2.1~eV$), there are 35 out of 28,468 detached cases (i.e., blue triangles at the top left quadrant of Figure~\ref{fig:forwad_highlight}(a)) mis-predicted as ``attached''. Similarly, 25 out of 29,187 attached cases (orange diamonds at the bottom right quadrant of Figure~\ref{fig:forwad_highlight}(a)) are mis-labeled ``detached''. Even counting in the other marginal cases, the mis-classification rate is lower than $0.2\%$. These mis-classified cases appears to congregate near the detachment onset point (i.e., all these cases have $T_{e,t}\in(1,4)~eV$). However, no common feature can be identified in terms of the control parameters $\x$.

Besides accuracy, speed is another important metric for a real-world application. Table~\ref{tab:speed} summarizes our predictive model's performance in terms of the wall-clock time required to predict diagnostic measurements based on controlled inputs. Even without any optimization, this model takes about $36~ms$ to carry out a prediction for one case - about 10,000 times faster than 1D UEDGE simulations which normally require a few minutes to find a converged solution. It is already within the minimum requirement for detachment control ($\sim100~ms$). The model efficiency also increases when predicting multiple cases (for $n<100,000$). 

\begin{table}
  \begin{center}
\def~{\hphantom{0}}
  \begin{tabular}{ccccccc}
       number of cases & 1 & 10 & 100 & 1,000 & 10,000 & 100,000 \\[3pt]
       speed ($ms$) & 36 & 37 & 42 & 51 & 178 & 2,259 \\ 
  \end{tabular}
  \caption{Wall-clock time v.s. number of cases.}
  \label{tab:speed}
  \end{center}
\end{table}


\section{Comparison with two-point models}\label{sec:comp}
Validation of the forward model for detachment prediction shows that our model is able to accurately predict UEDGE result with a dramatic speed up, but perhaps a more meaningful benchmark exercise would be comparing our newly developed model, as well as the UEDGE model with the widely used detachment prediction models nowadays, namely the analytical basic two-point model (2PM) \citep{stangeby2000plasma}, and the most sophisticated semi-analytical two-point model formatting (2PMF) \citep{stangeby2018basic}. 

\subsection{Basic two-point model}
\begin{figure}
  \centering
  \includegraphics[width=\textwidth]{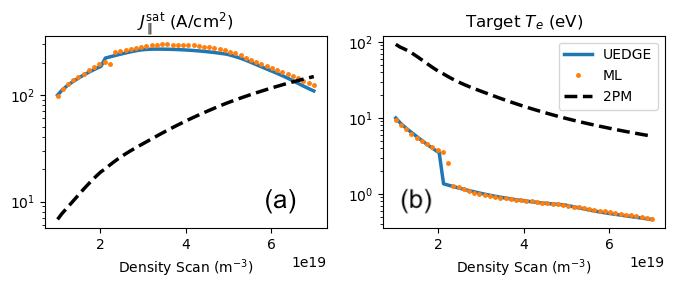}
  \caption{Upstream density scan of (a) $J_\parallel^\text{sat}$ and (b) $T_{e,u}$ for UEDGE (blue), machine learning data-driven model (orange) and basic two-point model(black dashed) with $P_\text{inj}=1.71~MW$, $f_Z=0$.}
\label{fig:2pm}
\end{figure}

The basic two-point model (2PM) is derived to evaluate $T_{e,u}$, $T_{e,t}$ and $n_{e,t}$ for given $n_{e,u}$, parallel heat flux $q_\parallel$ and flux tube length $L$ based on particle, pressure and power balance. Because of its simplicity, basic 2PM has been implemented in the tokamak detachment control as a priori to estimate the degree of detachment \citep{eldon2022enhancement}. Because 2PM doesn't account for momentum and power loss between the upstream and downstream points, we therefore benchmark UEDGE, the forward detachment model and the 2PM on a case with zero-impurity to remove the impurity radiation. Figure~\ref{fig:2pm} displays upstream density scan of $J_\parallel^\text{sat}$ and $T_{e,t}$ for the three models. Clearly, the data-driven model prediction and the UEDGE simulation results are in good agreement while the 2PM gives a quite different result.
This one order of magnitude discrepancy between our data-driven model/UEDGE and 2PM is due to the more comprehensive physics in our model/UEDGE. For instance, in this comparison the upstream ion temperature is much higher than the electron temperature in the upstream region in the UEDGE simulations due to slow ion parallel thermal conduction whereas the electron and ion temperatures are assumed to be the same in the basic 2PM. Additionally, in the basic 2PM, the parallel heat flux is assumed to be conductive using the Spitzer-H\"arm formula; while in 1D UEDGE simulations, the parallel heat flux is assumed to be local flux-limited thermal transport. UEDGE is only able to recover 2PM result by further simplifying simulation (e.g., using slab configuration and turn off non-ideal terms) and manually adjusting boundary conditions to match 2PM assumptions, such as $T_e=T_i$ at upstream.

\subsection{Two-point model formatting (2PMF)}
To address the volumetric momentum and power losses, and the magnetic geometry effects, the basic two-point model has been extended to semi-analytical two-point model formatting (2PMF) with pre-fitted power and momentum loss coefficients $f_\text{cooling}$, $f_\text{mom-loss}$ \citep{stangeby2018basic}. Benchmark between our data-driven model, UEDGE simulation and 2PMF are also performed. Two fitting curves (from \citep{stangeby2018basic}) are proposed for both power and momentum loss coefficients in 2PMF,

Stangeby fitting formula 1:
\begin{align}
  1-f_\text{cooling} &= [1-\exp({-T_{e,t}/2.4})]^{1.9}\\
  1-f_\text{mom-loss} &= [1-\exp({-T_{e,t}/0.8})]^{2.1}
\end{align}

\begin{figure}
  \centering
  \includegraphics[width=\textwidth]{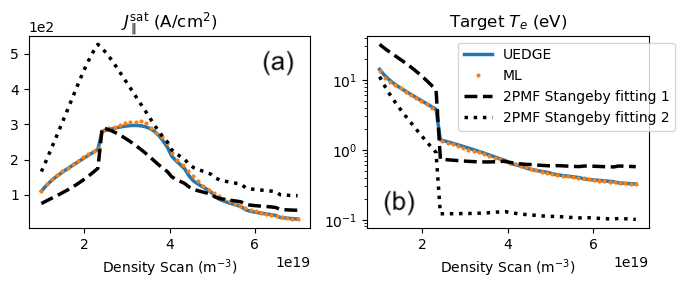}
  \caption{Upstream density scan of (a) $J_\parallel^\text{sat}$ and (b) $T_{e,u}$ for UEDGE (blue), machine learning data-driven model (orange) and two-point model formatting (black lines) with $P_\text{inj}=2.66~MW$, $f_Z=0.02$.}
\label{fig:2pmf}
\end{figure}

Stangeby fitting formula 2:
\begin{align}
  1-f_\text{cooling} &= 0.9[1-\exp({-T_{e,t}/6})]^{1.7}\\
  1-f_\text{mom-loss} &= 1.3[1-\exp({-T_{e,t}/1.8})]^{1.6}
\end{align}
where UEDGE observed $T_{e,t}$ is used to estimate $f_\text{cooling}$ and $f_\text{mom-loss}$. Figure~\ref{fig:2pmf} displays upstream density scan of $J_\parallel^\text{sat}$ and $T_{e,t}$ for the three models. The results from 2PMF compare with UEDGE simulation results better than those from the basic 2PM. All three models show two common features: (1) the ion saturation current density rollover and (2) temperature cliff at the onset of detachment. Once again, our data-driven model prediction matches UEDGE simulation results very well. Although there is no qualitative disagreement between our data-driven model/UEDGE and 2PMF, quantitatively the predictions between data-driven model/UEDGE and 2PMF can be off by an order of magnitude. This is because 2PMF prediction depends heavily on the choice of fitting curves for the power and momentum loss coefficients which may simplify the nonlinear dynamics setting the divertor plasma conditions. As illustrated in Figure~\ref{fig:2pmf}, fitting formula 1 does a better overall job than fitting formula 2 for this test case as $T_{e,t}$ predicted by fitting formula 2 is near an order of magnitude lower than UEDGE result when divertor plasma is detached ($n_{e,u}>2.4\times 10^{19}~m^{-3}$).


\section{Model applications}\label{sec:app}

As mentioned in the introduction, the main motivation of developing such a surrogate model is to enable reliable and fast detachment prediction for integrated machine design, scenario development and real time plasma control. Heat exhaust is not a severe issue for current tokamaks due to the overall limited power output. Therefore, the scrape-off layer (SOL) and divertor plasma dynamics is less of a concern and most of the effort so far has been focused on exploring operation scenarios with improved fusion performance, equilibria, and plasma stability control inside the separatrix, e.g., core-edge integration. However, for future high fusion power devices such as reactors, their operation space must also fulfill constraints posed by divertor’s material and engineering limits. Likewise, fusion burn control needs to incorporate divertor heat and particle exhaust solutions such as detachment. In other words, designing and operating future devices require core-edge-SOL/divertor integration. SOL/divertor modelling suffers from the accuracy-speed trade-off similar to many other research. Current SOL/divertor transport codes such as UEDGE and SOLPS, are too slow for these applications as they are designed for physics investigations; while the basic two-point model and two-point model formatting are fast but perhaps over-simplify the problem. Our data-driven model overcomes this accuracy-speed gap with some room to prioritize one factor over another depending on the application. For instance, speed is the top criterion for plasma control (either in a simulator or in the actual plasma control system). Real time or even fast than real time prediction is required in order to activate actuators in time. On the other hand, accuracy is likely weighted more than speed for device design and scenario development.

The accuracy of the proposed data-driven approach relies on the quality and quantity of the training data set. We would like to remark that tokamak edge plasma contains very rich physics with many factors have influence on the detachment onset or threshold, such as 2D/3D effects, divertor plate geometry, multi-species and/or multi-charge-state impurity, wall condition, and so on. Because of the 1D flux-tube mesh simplification, 1D UEDGE simulations under-estimate the detachment onset temperature $T_{e,t}$ and over-estimate the peak radiation amplitude, our current model no surprisingly picks up these unfavorable predictions. One would expect that the performance of model improves in terms of matching real experimental measurement when the training data is extended to incorporate richer physics in a more realistic experiment setting, i.e., once trained upon higher quality data sets either from higher fidelity numerical models such as 2D UEDGE/SOLPS-ITER or experiments. Even though the underlying methodology will be the same, the architecture of neural networks likely needs to be modified. Notably, synthetic, or real diagnostic measurements will have a complicated format with 2D simulations or experiments. In a realistic tokamak, both Langmuir probe and Thompson scattering are multi-channel; therefore $J_\parallel^\text{sat}$, $T_{e,u}$, $n_{e,t}$ and $T_{e,t}$ become sparse 1D arrays in space (point data), spectroscopic and bolometer diagnostics provide radiation power/strength at certain wavelength or over the entire spectrum (e.g., 1D volume averaged data), visible and IR cameras give image (i.e., 2D data with projected volume and certain range of wavelength averaged quantity). Moreover, there are likely some discrepancies between model produced synthetic diagnostic data and real experimental measurements due to various reasons such as model simplification, inherent instrumentation noise and so on. Handling these multi-modal diagnostics together in a consistent manner, as well as bridging the gap between simulation and experimental data will be addressed in our future study.

Even though the data-driven detachment prediction model presented here is based on 1D UEDGE simulations, it could be readily integrated for detachment control application. At first glance, current model is similar to the analytical basic two-point model which has recently been implemented in KSTAR tokamak detachment control system~\citep{eldon2022enhancement} as both models are meant to predict divertor or downstream plasma state. Comparing to the basic two-point model, our model features additional detachment front prediction. Additionally, due to fewer simplifications made in UEDGE model such as $B$ variation, local flux-limited thermal transport, impurity radiation effects, etc, our model should give more reliable predictions than two-point model in certain circumstances, resulting improved control performance of detachment and overall plasma confinement.

\section{Summary}\label{sec:conc}
In this paper, we explore a new physics model-based approach to predict divertor detachment by leveraging the ``latent feature space'' concept in machine learning research. 
As a proof of concept study, highly efficient 1D UEDGE model which contains the crucial physics ingredients of detachment are used to simulate the plasma and neutrals along the open magnetic field lines in the scrape-off layer (SOL) and to generate our training and validation data sets. Over 160,000 simulations with three varying UEDGE model inputs $\x$ (e.g., different upstream density, injection power and carbon fraction) are performed to cover the normal DIII-D tokamak operation parameter region; and five synthetic diagnostic measurements such as upstream temperature, electron density, temperature and saturation current at divertor target, as well as radiation profile are collected as the diagnostic set $\y$. The latent space as well as the latent space representation of plasma state $\z$ are then identified by compressing $\y$ through an autoencoder. Sequentially, a forward surrogate model is trained to make predictions of $\z$ from UEDGE model inputs $\x$; then the trained decoder is used to reconstruct diagnostic measurement $\y$ back in configuration space. 
We find that a 6-dimensional latent space is good enough to closely yield a match for the true system in configuration space (i.e., the synthetic diagnostic measurements), and the forward detachment prediction model ($\x\to\z\to\y$) also produces quite accurate predictions (relative error on the order of a few percent statistically) with at least $10^4$ speed-up comparing to the UEDGE simulations. 

Our pilot study demonstrates that the complicated divertor/SOL plasma state has a low-dimensional representation in latent space. Therefore, this new latent space description of plasma state not only can be used to construct a fast and robust surrogate model for steady-state detachment prediction as revealed in this paper, but also has the potential to be used for dynamical control once the critical plasma nonlinear dynamics (in latent space) was successfully identified.

\section*{Acknowledgement}
The authors thank the anonymous reviewers for their comments and suggestions.

\section*{Funding}
This work is performed under the auspices of the U.S. Department of Energy (US DOE) by Lawrence Livermore National Laboratory under Contract DE-AC52-07NA27344 through the pilot project ``A critical step for development of machine learning surrogate models for tokamak divertor-plasma detachment control". LLNL-JRNL-836030.

\section*{Declaration of interests}
The authors report no conflict of interest.

\section*{Data availability statement}
The data that support the findings of this study are available from the corresponding author upon reasonable request.


\bibliographystyle{jpp}

\bibliography{sdetach}

\end{document}